\documentclass[prc,twocolumn,preprintnumbers,showpacs,superscriptaddress]{revtex4}
\usepackage[colorlinks=true,linkcolor=blue,citecolor=red,urlcolor=darkred]{hyperref}
\usepackage{gensymb}
\usepackage{amsmath,amssymb,epsfig,bm,mathrsfs,color}
\usepackage{natbib}
\usepackage{slashed}
\usepackage{color}
\newcommand{\be}{\begin{equation}}
\newcommand{\ee}{\end{equation}}
\newcommand{\bea}{\begin{eqnarray}}
\newcommand{\eea}{\end{eqnarray}}

\newcommand{\mnras}{Mon. Not. R. A. S.}

\newcommand{\vecA}{\bm A}

\newcommand{\vecB}{\bm B}

\usepackage[normalem]{ulem}  

\definecolor{darkred}{rgb}{0.7,.1,.2}

\preprint{INT-PUB-24-031}
\begin{document}

\title{Josephson currents in neutron stars}

\author{Armen Sedrakian}%
\affiliation{Frankfurt Institute for Advanced Studies,
D-60438 Frankfurt am Main, Germany}
\affiliation{ Institute of Theoretical Physics,
University of Wroclaw, 50-204 Wroclaw, Poland}  
\email{sedrakian@fias.uni-frankfurt.de}

\author{Peter B. Rau}
 \affiliation{Institute for Nuclear Theory, University of Washington, Seattle, Washington, 98195-4550 USA}
 \altaffiliation{Present address: Columbia Astrophysics Laboratory, Columbia University, New York, New York 10027 USA}
 \email{peter.rau@columbia.edu}

\date{January 3, 2025}

\begin{abstract} {We demonstrate that the interface between $S$-wave
    and $P$-wave paired superfluids in neutron stars induces a neutron
    supercurrent, a charge-neutral analog of the Josephson junction
    effect in electronic superconductors. The proton supercurrent
    entrainment by the neutron superfluid generates, in addition to
    the neutral supercurrent, a charged current across the
    interface. Beyond this stationary effect, the motion of the
    neutron vortex lines responding to secular changes in the neutron
    star's rotation rate induces a time-dependent oscillating
    Josephson current across this interface when proton flux tubes are
    dragged along with them. We show that such motion produces
    radiation from the interface once clusters of proton flux tubes
    intersect the interface. The power of radiation exceeds by orders
    of magnitude the Ohmic dissipation of currents in neutron
    stars. This effect appears to be phenomenologically significant
    enough to heat the star and alter its cooling rate during the
    photon cooling era.  }
\end{abstract}

\maketitle

\section{Introduction}
\begin{figure}[t] 
\begin{center}
\includegraphics[width=1.2\linewidth,keepaspectratio]{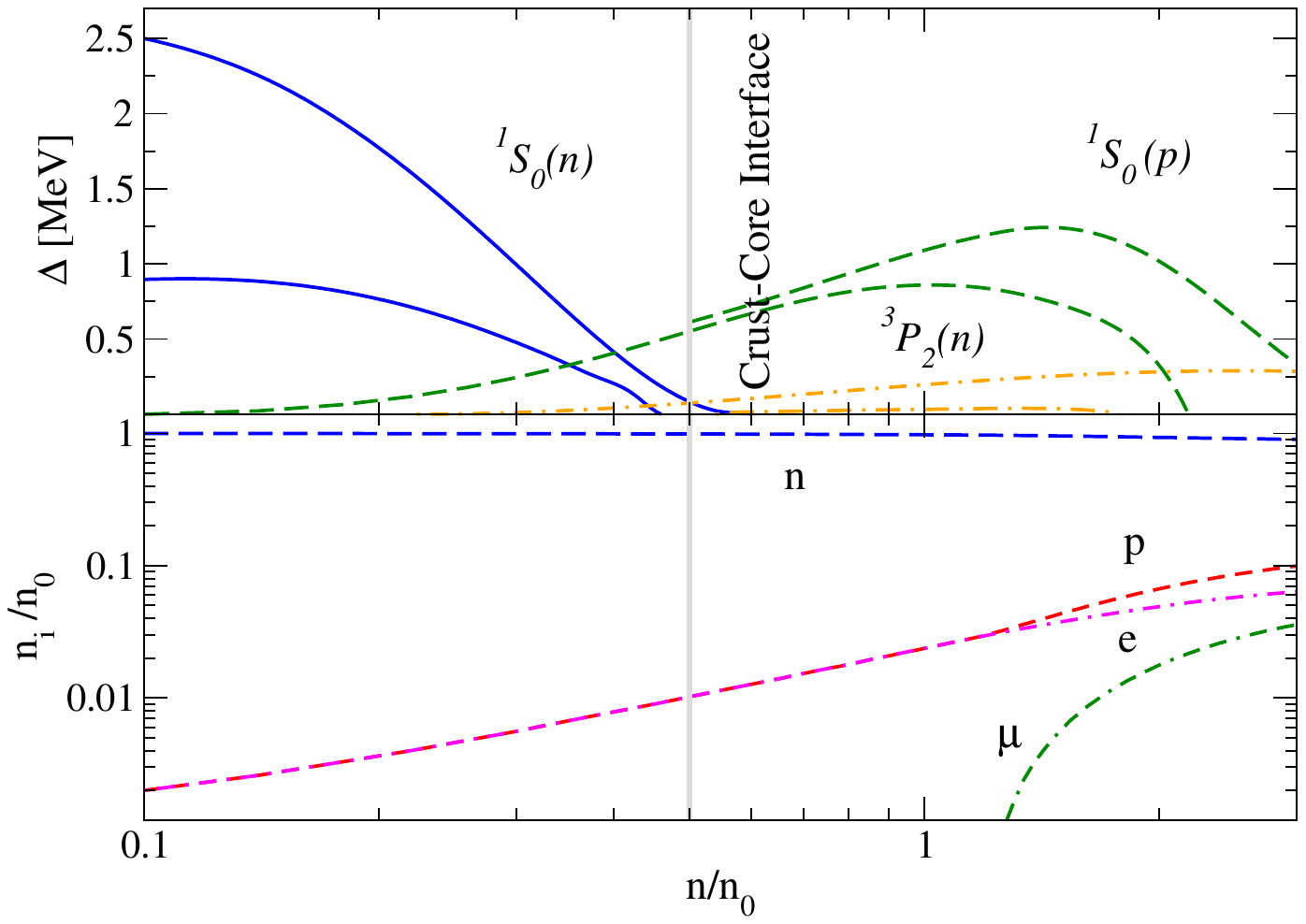}
\caption{Upper panel: dependence of pairing gaps for neutrons in
  $^1S_0$ (solid lines) and $^3P_2$--$^3F_2$ (dot-dashed lines) and
  for protons in $^1S_0$ (dashed lines) on baryonic density in units
  of nuclear saturation density. The upper and lower lines correspond
  to lower and upper bounds as discussed in
  Ref.~\cite{Sedrakian2019EPJA}.  Lower panel: composition of the core
  of a neutron star at $T=0$ used to obtain the gaps. The vertical
  line shows the crust-core interface at the density $n/n_0=0.5$.}
\label{fig:gaps} 
\end{center}
\end{figure}

Neutron stars are unique objects where quantum physics manifests
itself on stellar-size macroscopic scales as their rotational dynamics
and magnetic fields are supported by quantized vortex networks
extending over macroscopic scales; for reviews,
see Refs.~\cite{Graber2017,Haskell2017}. At low densities, their inner crust
contains a neutron superfluid, which forms spin-zero Cooper pairs due
to nuclear attraction in the $^1S_0$ partial wave channel.  At high
densities, corresponding to the core of the star, the dominant
attractive partial wave channel is the $^3P_2$-$^3F_2$; i.e., neutrons
are paired in a spin-one state~\cite{Sedrakian2019EPJA,Chamel2024}. An
alternative $P$-wave channel at high density is the $^3P_0$ channel if
the $^3P_2$-$^3F_2$ pairing is suppressed by the spin-orbit
interaction ~\cite{Krotscheck2023ApJ}. The density at which crust-core transition occurs is given by the density
$n^{\sharp}\simeq 0.5n_0=0.08$ fm$^{-3}$, which is defined by the
dissolution of protons bound in nuclei into a continuum, forming a
fluid~\cite{Grams2022}. As seen from Fig.~\ref{fig:gaps}, the
transition from $S$- to $P$-wave neutron superfluid occurs at
about the same density.
Unbound protons at and above $0.5n_0$ pair in the spin-zero $S$-state
and are admixed with the $P$-wave neutron superfluid. The transition
between $S$ and $P$-wave superfluids can occur either at a sharp
interface or via a formation of a mixture of these phases 
within some density domain. We will assume a sharp interface between
the two superfluids, but we will briefly comment on the implications
of a mixed phase.

The focus of this work is on the interface between two {\it neutral}
$S$-wave and $P$-wave superfluids. We show that a current emerges at
this interface, in analogy to Josephson effects in metallic
superconductors~\cite{Landau1980,Ketterson2016}, due to the difference
in the phases of these superfluids. Specifically, we show the following
stationary case:
\begin{description}
\item (a) a neutron superflow arises at the $S$-wave/$P$-wave
  superfluid interface due to the difference in the phases of these
  superfluids, i.e., the interface acts as a Josephson junction. An
  analogous effect in neutral superfluid ultracold atomic gases has
  been observed recently~\cite{Pezze2024NatCo}.
\item (b) Because of the entrainment of the protons by the motion of
  the neutrons~\cite{Sedrakian1980,Alpar1984,Allard2021}, the neutron
  superflow induces a charged current through the interface as well.
\item (c) Going beyond the stationary limit, we show that  a type
  of non-equilibrium Josephson effect arises because the star changes
  its spin frequency on secular timescales.
\end{description}

Neutron superfluid rotates by
forming an array of quantum vortices and the proton superconductor is
type II at relevant densities and sustains flux tubes that carry the
magnetic flux of the star within the superconducting
regions~\cite{Glampedakis2011,Gusakov2016a,Graber2017,Haskell2017,Rau2020}.
The secular deceleration of neutron stars due to (primarily) their
magnetic dipole radiation leads to an expansion of the neutron vortex
lattice and neutron vortex lines (and eventually proton flux tubes)
crossing the interface.  This time-dependent process leads to the current
oscillations through the interface---an analog of the time-dependent
Josephson effect.  The dynamics of a coupled network of neutron vortex
and proton flux tube arrays is not well understood over timescale
characteristic for glitches and their relaxation due to their
complexity. Below, we assume that the flux tubes comove (on secular
time scales) with neutron vortices and show that the crust-core
interface radiates electromagnetically. The amount of radiation
emitted by the flux tubes is significant enough to deposit
phenomenologically important amounts of heat at the crust-core
interface, thereby altering the cooling behavior of neutron stars.  We
do not address the problem of glitches, which can be associated with
the interfaces between various phases, in particular superfluids; see,
for example, Refs.~\cite{Sedrakian1999, Marmorini:2020zfp, Poli2023}.

\section{Stationary currents across the interface}

\begin{figure}[t] 
\begin{center}
\includegraphics[width=1\linewidth,keepaspectratio]{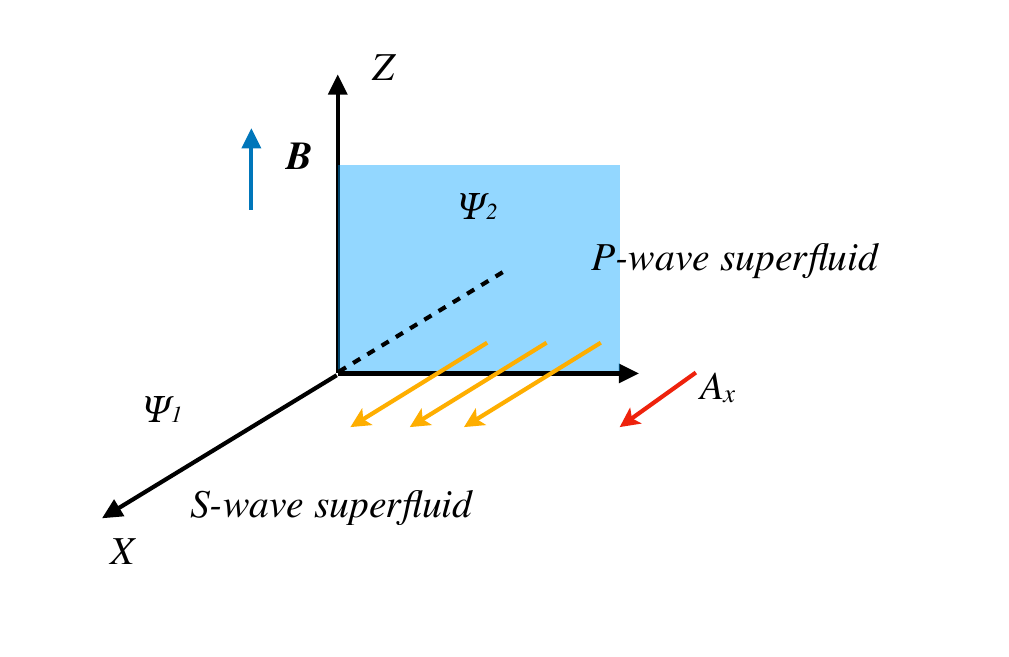}
\caption{Illustration of the local setup to compute the Josephson
  current. The interface, separating the superfluids, is the $yz$
  plane with $x>0$ corresponding to the $S$-wave condensate and $x<0$
  to the $P$-wave condensate.  The Josephson current (yellow arrows)
  flows across the interface parallel to the $x$-axis. In the case of
  proton flux tubes the magnetic field $\vecB$ is locally along the
  $z$-axis and the vector potential $\vecA$ varies along the $x$-axis,
  while its constant $y$-component does not contribute to its curl.}
\label{fig:JC_fig1} 
\end{center}
\end{figure}

Consider a setup where the interface is the $yz$ plane of a Cartesian
coordinate system. The $x> 0$ region corresponds to $S$-wave neutron
($n$) superfluid characterized by an equilibrium Ginzburg-Landau (GL)
order parameter
(wave function) $\psi_{1,n}^{\infty}$ and the $x<0$ region corresponds
to $P$-wave superfluids with GL wave function
$\psi_{2,n}^{-\infty}$. This is illustrated in
Fig.~\ref{fig:JC_fig1}. A $P$-wave superfluid has a more complex order
parameter~\cite{Muzikar1980,Masuda2016,Masaki2020,Yasui2020,Masaki2022,Kobayashi2022,Kobayashi2023}
than the scalar one adopted here, but the spin dynamics of the Cooper
pairs is not essential for our discussion.

The equations of motion for the wave functions $\psi_{p/n}$ of proton
($p$) and neutron $(n)$ condensates (i.e. the first GL equations) are
derived in Appendix~\ref{app:A} and are given by
\bea\label{eq:GL_p_1}
&&\hspace{-0.5cm} a_p \psi_p+b_p\left|\psi_p\right|^2 \psi_p+g_{np}\left|\psi_n\right|^2
\psi_p\nonumber\\
&&\hspace{2cm}+\frac{1}{4m_p}\left(-i\hbar \bm{\nabla} -\frac{2 e}{c} \bm{A}\right)^2 \psi_p=0,\\
\label{eq:GL_n_1}
&&\hspace{-0.5cm} a_n\psi_n+b_n\left|\psi_n\right|^2 \psi_n+g_{np}\left|\psi_p\right|^2
\psi_n - \frac{\hbar^2}{4m_n}\nabla^2 \psi_n=0,
\eea
where $a_{n,p}< 0$ in the superfluid state, $b_{n,p}>0$, $\bm{A}$ is
the vector potential, $e>0$ is the unit charge, $g_{np}$ describes the coupling between the amplitudes of the
condensates, $c$ is the speed of light and $m_{n/p}$ are the bare masses
of the neutron/proton. The boundary condition at the interface between the
superfluids is a generalization of the ordinary GL boundary conditions
to account for the tunneling of the pairs across the
boundary~\citep{Ketterson2016}
\be \label{eq:boundary}\bm n\cdot \left( \bm \nabla -
  \frac{2ie}{\hbar c}\bm A\right)
\psi_{1,2,j}\Big\vert_{\rm Boundary} = \pm \frac{\psi_{2,1,j}\vert_{\rm Boundary}}{\zeta_{j}}, 
\ee
where $j = n,p$, the coefficient $\zeta_{n/p}^{-1}$ is proportional to the
permeability of the barrier between the superfluids~\cite{Landau1980}
and $\bm{n}$ is the unit vector orthogonal to the boundary.
The proton charge current density and neutron mass current density are obtained by variation of the GL
functional with respect to the vector potential $\bm{A}$ and neutron
velocity $\bm{v}_n$, respectively:
\begin{align}
\bm{j}_p=&
\frac{e\hbar}{2im_p}\left(1-\frac{\rho_{pn}}{\rho_{pp}}\right)\left(\psi_p^*\bm{\nabla}\psi_p-\psi_p\bm{\nabla}\psi_p^*\right)
\nonumber\\
{}&+\frac{e\hbar}{2im_p}\frac{\rho_{pn}}{\rho_{nn}}\left(\psi_n^*\bm{\nabla}\psi_n-\psi_n\bm{\nabla}\psi_n^*\right)
\nonumber\\
{}&-\frac{e^2}{m_p^2c}(\rho_{pp}-\rho_{pn})\bm{A},
\label{eq:GL_p_2}
\\
\bm{j}_n=&\frac{\hbar}{2i}\left(1-\frac{\rho_{pn}}{\rho_{nn}}\right)\left(\psi_n^*\bm{\nabla}\psi_n-\psi_n\bm{\nabla}\psi_n^*\right)
\nonumber\\
{}&+\frac{\hbar}{2i}\frac{\rho_{pn}}{\rho_{pp}}\left(\psi_p^*\bm{\nabla}\psi_p-\psi_p\bm{\nabla}\psi_p^*\right)
-\frac{e}{m_pc}\rho_{pn}\bm{A}.
\label{eq:GL_n_2}
\nonumber\\
\end{align}
The derivation of these equations and definition of the elements of the density matrix $\rho_{ij}$, $i,j\in p,n$ are
given in Appendix~\ref{app:A}. The off-diagonal term $\rho_{pn}=\rho_{np}$
describes the kinetic coupling between the two condensates known as
\textit{entrainment}. The combination of \eqref{eq:boundary} with \eqref{eq:GL_p_2} and
\eqref{eq:GL_n_2}, and the requirement of invariance upon a gauge
transformation $\bm{A}\rightarrow\bm{A}+\bm{\nabla}\alpha$
for a scalar function $\alpha$ give the Josephson currents across the interface
\begin{align}
 \bm{n} \cdot \bm{j}_p=&\frac{e \hbar}{ m_p\zeta_p }
 \frac{\left(\rho_{p p}-\rho_{p n}\right)}{\rho_{pp}}
 \left|\psi_{1, p}\right|\left|\psi_{2, p}\right| \sin \left(\Delta\theta_{p}\right)
 \nonumber \\
 &+\frac{e \hbar }{m_p \zeta_n}  \frac{\rho_{p n}
 }{\rho_{nn}}
 \left|\psi_{1, n}\right|\left|\psi_{2, n}\right| \sin \left(\Delta\theta_{n}\right).
 \label{eq:p_current}
 \\
\bm{n} \cdot \bm{j}_n=&\frac{\hbar}{\zeta_n}
\frac{\rho_{n n}-\rho_{p    n}}{\rho_{n n}}
\left|\psi_{1, n}\right|\left|\psi_{2, n}\right| \sin \left(\Delta\theta_{n}\right)
\nonumber\\
&+\frac{\hbar}{\zeta_p} \frac{\rho_{p n}}{\rho_{p p}}
\left|\psi_{1, p}\right|\left|\psi_{2, p}\right| \sin \left(\Delta\theta_{p}\right),
\label{eq:n_current}
  \end{align}
 where we used $\psi_j=\left|\psi_j\right| e^{i \theta_j}$ for $j=n,p$ and where the gauge invariant phase difference is
\bea\label{eq:Delta_theta}
  \Delta\theta_{i} = \theta_{2, i}-\theta_{1, i} -\frac{2\pi}{\Phi_0}\delta_{i,p} \int_p
  \bm{A} \cdot d \bm{l}.
  \eea
  Here $\Phi_0=\pi \hbar c/e$ is the flux quantum for the 
  proton ($p$) condensate and $\delta_{ip}$ is a Kronecker delta. 

  Equations~\eqref{eq:p_current} and~\eqref{eq:n_current} demonstrate
  that a stationary supercurrent flows across the interface between
  the $S$- and $P$-wave superfluids, driven by the phase difference
  $\Delta\theta_n$ of the two superfluids. This phase difference
  arises from a change in the neutron pairing interaction,
  transitioning from $^1S_0$ to $^3P_2$--$^3F_2$-wave as the neutron
  density increases, as shown in Fig.~\ref{fig:gaps}. Additionally,
  Eq.~\eqref{eq:p_current} reveals that the entrainment effect results
  in a proton electric supercurrent across the interface, also induced
  by the phase difference $\Delta\theta_n$.

  If uncoupled from the neutron condensate, the proton condensate
  order parameter is the same on both sides of the interface
  $\psi_{1,p}= \psi_{2,p}$ and corresponds to $^1S_0$ pairing, which
  is due to the lower density of protons compared to neutrons under
  $\beta$-equilibrium in neutron stars. However, because of the
  entrainment between the neutron and proton condensate wave
  functions, $\psi_{1,p}$ and $\psi_{2,p}$ are no longer required to
  be equal if $\psi_{1,n}\neq\psi_{2,n}$.  A density jump at the
  $S$–$P$ transition interface, either for neutrons or protons, is
  energetically implausible. Consequently, the magnitudes of their
  order parameters must remain continuous. Thus, the phase difference
  in the neutron condensate necessitates a corresponding phase
  difference in the proton condensate.  This can be seen by examining
  the Ginzburg-Landau free energy, which should be equal on both sides
  of the junction since this is the condition for the transition from
  $S$- to $P$-wave neutron pairing. If the condensate densities are
  the same on both sides of the junction $|\psi_{n,1}|=|\psi_{n,2}|$
  and $|\psi_{p,1}|=|\psi_{p,2}|$, the only contribution to the
  Ginzburg-Landau free energy which can be different across the
  junction is the kinetic part, given by (see Appendix~\ref{app:A})
 \begin{align}
F_{\rm kin}={}&\frac{\hbar^2}{8 m_p^2}\left(\rho_{p p}-\rho_{p
    n}\right)\left(\bm{\nabla} \theta_p-\frac{2 e}{\hbar c}
  \bm{A}\right)^2
  \nonumber\\
  &{}+\frac{\hbar^2}{8 m_n^2}\left(\rho_{n n}-\rho_{p
    n}\right)\left(\bm{\nabla} \theta_n\right)^2
    \nonumber\\
&{}+\frac{\hbar^2
  \rho_{p n}}{4 m_p m_n}\left(\bm{\nabla} \theta_p-\frac{2 e}{\hbar c}
  \bm{A}\right) \cdot \bm{\nabla} \theta_n.
  \label{eq:F_kinMain}
 \end{align}

 We can evaluate $F_{\rm kin}$ on either side of the junction and set
 both sides equal to each other. For this equality to be true while
 the neutron phase is different across the junction
 $\theta_{1,n}\neq\theta_{2,n}$, either (a) the gradients in the
 neutron phases must be equal
 $|\bm{\nabla}\theta_{1,n}|=|\bm{\nabla}\theta_{2,n}|$ or (b) the
 proton phase (or more specifically its gradient) must also be
 different across the junction. This leaves the possibility of
 $\theta_{1,p}\neq\theta_{2,p}$.


\section{Time-dependent currents across the interface}

Neutron stars decelerate under the action of external braking torques,
which leads to a decrease in the average number of quantum neutron
vortices in the superfluid. By continuity, the number of neutrons
vortices is reduced through their motion away from the rotation
axis. Because of the entertainment, neutron vortices carry a
fractional flux localized close to their cores. In addition, the
stellar magnetic field in the type II superconducting core exists in
the form of proton flux tubes, which will be rearranged by such
motion. The average number of proton flux tubes intersecting the area
occupied by a single neutron vortex is very large, of the order of
$n_{\Phi}/n_V\sim 10^{13}$, where the number densities of proton flux
tubes and neutron vortices are given by $n_{\Phi}=\bar B/\Phi_0$, and
$n_V=2\Omega/\kappa$ with quantum circulation
$\kappa=\pi \hbar /m_n^*$ with $m_n^*$ being the neutron (effective)
mass. Here $\Omega$ is the rotation frequency and $\bar B$ is the
average magnetic field. The field may be inhomogeneously distributed
across a single neutron vortex. For, example, the vortex-cluster model
envisions the clustering of proton flux tubes around the neutron
vortex~\cite{Sedrakian1995ApJ}. The interaction between neutron vortex
arrays and flux tubes in a type II superconductor is generally
complex. This complexity arises from the unknown geometry of the flux
tube arrays, which is dictated by the global magnetic field
configuration of the star (e.g., poloidal versus toroidal), as well as
the uncertain strength of the interactions between vortices and flux
tubes. As limiting cases, one can envision two scenarios: in response
to the star's deceleration and the motion of neutron vortices, the
flux tubes could either move together with the vortices or remain
anchored to the crustal magnetic field, staying motionless. Since our
focus is on the long-term dynamics of the vortex–flux tube
conglomerate over secular timescales, which are much longer than the
typical timescales associated with glitches and postglitch
relaxations, we will assume that the flux tubes, when averaged over
long timescales, move at the same speed as the neutron vortex array.

The motion of neutron vortex lines and proton flux tubes outward from
the core of a neutron star will serve as a source of time-dependent
Josephson currents across the $S$--$P$ neutron superfluid phase
transition interface. Such a current requires a time-varying phase difference across the junction. A neutron vortex line passing through
the junction will induce a time-dependent phase difference in the neutron
superfluid. This is because in the core of a neutron vortex, the
superfluidity is suppressed and outside this core, the phase of the
order parameter depends on the distance from the axis of the vortex
$\varrho$ according to
$\bm{\nabla} \theta_n=(1/{\varrho})~ \hat{\varphi}$ (see
Eqs.~\eqref{eq:velocities} and \eqref{eq:vortex_v_n}) and thus changes
logarithmically as $\theta_n(\varrho) = \ln\,\varrho$ with the
distance between the vortex and the interface,  where $\varrho$ and
$\varphi$ are the cylindrical radius and angle in the plane orthogonal
to the vortex circulation.  

If a neutron vortex is moving through the interface then the neutron order parameter will acquire a time-dependent phase
\begin{equation}
    \Delta\theta_{n}=\theta_{1,n}-\theta_{2,n}+\omega_V t,
\end{equation}
where $\theta_{1,n}$ and $\theta_{2,n}$ are the static phases on
either side of the junction and $\omega_V t$ is a time-dependent part
induced by the motion of a neutron vortex line through the junction
with period $2\pi/\omega_V$, with $\omega_V = v_{Lr}/d_n$, where
$v_{Lr}$ is the radial velocity of neutron vortex lines averaged over
secular time scales and is given by 
\bea
v_{Lr} = -\frac{\dot \Omega}{2\Omega}r,
\label{eq:VortexLineSpeed}
\eea
where $r$ is the cylindrical radius and $d_n$ is the intervortex
spacing. Equation~(\ref{eq:VortexLineSpeed}) follows from the conservation
of vortex number density $n_V$. Because of superfluid entrainment, the
time-dependent neutron phase induces a time-dependent proton current
through the junction given by
\bea 
j_{V}\equiv \bm{n}\cdot\bm{j}_p(t)=j_{0,pn}\sin{(\theta_{2,n}-\theta_{1,n}+\omega_V t)},
\label{eq:NeutronPhaseDiffCurrent}
\eea
where 
\bea 
j_{0,pn}\equiv\frac{e\hbar}{m_p\zeta_n}\frac{\rho_{pn}}{\rho_{pp}}|\psi_{1,n}||\psi_{2,n}|.
\eea
As we showed in the previous section, a time-dependent neutron phase
difference can induce a time-dependent proton phase difference across
the interface. Additionally, the motion of proton flux tubes through
the junction will induce a time-dependent vector potential within the
interface, so that the resulting gauge-invariant phase difference will be
time dependent. Consider now a single flux tube crossing the interface
between the two superfluids, moving along with the vortex lines. The
tube is assumed to be locally straight and along the $z$ axis of our
setup and the current is flowing along the $x$ axis. The variations of
the vector potential are along the $x$ axis so that the nonvanishing
component is $A_x (x)= -H_z(x) y$ (see Fig.~\ref{fig:JC_fig1}).
For this setup
Eq.~\eqref{eq:Delta_theta} for the protons can be written 
\bea\label{eq:Delta_theta_2} \Delta\theta_p
= \theta_{2,p} - \theta_{1,p} - \frac{4\pi yH_{0z}\lambda}{\Phi_0},
\eea
where $\lambda$ is the magnetic field penetration depth.  To evaluate
the integral in Eq.~\eqref{eq:Delta_theta}, we integrate between two
points deep in the $S$- and $P$-wave superfluids, respectively, and
use that the magnetic field of a flux tube is given as
$H_{z}(x)=H_{0z}\exp(-x/\lambda)$, where we used the large-distance
asymptotic form of the modified Bessel function $K_0$.  The passage of
a flux tube through the interface induces a time-dependent flux inside
the junction. For a proton flux tube, this has a frequency
$\omega_p = v_{Lr}/d_p$, where $v_{Lr}$ is given by Eq.~(\ref{eq:VortexLineSpeed}) 
since the neutron vortex lines transport the proton flux tube, and $d_p$ is the
interflux tube spacing. The maximal value of the flux
$\Phi_{\rm max}$ will correspond to the case where the flux tube
coincides with the interface, which has a width of roughly $2\xi_n$,
where $\xi_n$ is of the order of the coherence length of the neutron
superfluid. Therefore, the time-dependent flux can be written as as
\bea
\Phi_*(t) = \Phi_{\rm max} \cos(\omega_p t),\quad \Phi_{\rm max}   =8\pi \xi H_{0z}\lambda.
\eea

Thus, the time-dependent part of the Josephson current
\eqref{eq:p_current} now arises from the proton phase difference and is given by
\bea \label{eq:p_current_t-dependent}
j_{\Phi}= \bm{n} \cdot \bm{j}_p(t) &=&
j_{0,pp} \sin \left[\theta_{2,p}-\theta_{1,p} + \phi(t)\right],
\eea
where
\bea
j_{0,pp} \equiv
\frac{e \hbar}{ m_p\zeta_p }
\frac{ \left(\rho_{p p}-\rho_{p
     n}\right)}{\rho_{pp}}
\left|\psi_{1, p}\right|\left|\psi_{2, p}\right|,
\quad \phi(t) = \frac{\Phi_*(t)}{\Phi_0}.
 \eea
The passage of a flux tube array through the
interface thus induces an oscillating Josephson current in the
interface. Due to the entrainment effect, the neutron vortex lines are also magnetized, 
and so a neutron vortex passing through the junction will also induce a time-dependent phase difference in the proton condensate. However, the quantized flux within a neutron vortex is reduced compared to its value in a proton flux tube, and since the number density of proton flux tubes is much greater than of neutron vortex lines $n_{\Phi}\gg n_{V}$, we neglect the magnetization of the latter here.

\section{Radiation from the interface of two superfluids}

A time-oscillating Josephson current will lead to radiation of energy from the
interface. The time-average power radiated by an oscillating current
is given by 
\bea
\label{eq:rad_formula}
\langle P\rangle  = 
\frac{2\langle \dot{I}^2\rangle d^2}{3c^3},
\label{eq:DipoleRadPower}
\eea
where $\langle \dots \rangle$ denotes a time average, 
$I$ is the current, and $d$ is the length scale over which the current oscillates.

\begin{figure}[t] 
\begin{center}
\includegraphics[width=1\linewidth,keepaspectratio]{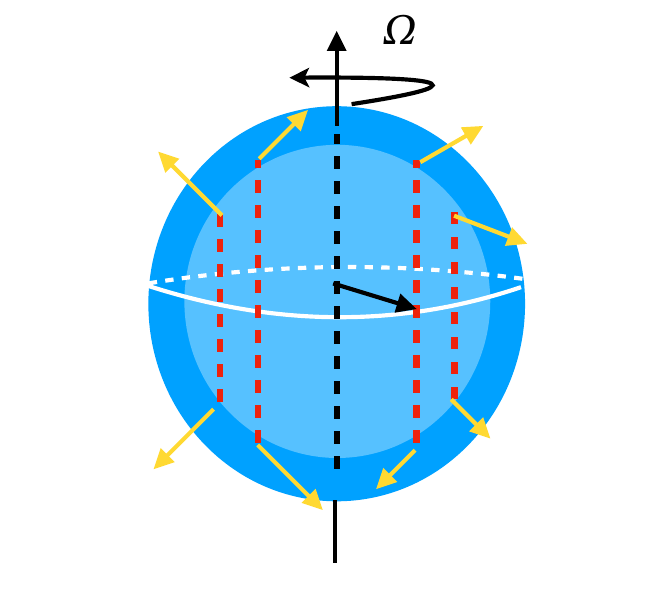}
\caption{Illustration of the computation of radiation (yellow arrows)
  at the interface between $S$- and $P$-wave superfluids. A collection of
  vortices or flux tubes (red lines) cross the
  interface. The effective macroscopic length over which 
  radiation is taking place is of the order of $R$.}
\label{fig:JC_fig2} 
\end{center}
\end{figure}

To obtain the power radiated over the entire star requires specifying
the geometry of the problem. We assume that the spin axis is along the
$z$ direction of the cylindrical or polar direction of the spherical
coordinate system, the former dictated by the geometry of the neutron
vortex lattice and the latter by the spherical shape of the star, see
Fig.~\ref{fig:JC_fig2}.  For the sake of simplicity, we assume that
the interface between the superfluids is a cylinder coaxial with the
rotation axis, with a height $\ell_z\le R$ centered at the equator.
We will associate each neutron vortex with a cluster of proton flux tubes,
as indicated in the previous section.  We note that in spherical geometry only a
segment of a flux, at any given time, crosses the interface; therefore,
the approximation $\ell_z\le R$ might appear unjustified. However,
consider moving from the North Pole of the star toward the equator
along a meridian. As one proceeds, the path will
intersect with vortices and associated flux tubes, which are spaced at
microscopic distances from each other. Therefore, the entire meridian
will contribute to radiation, even though at any given point, a
different flux tube will be radiating.  

We begin by estimating the radiation from a single neutron vortex
lines passing through a junction. To obtain the power radiated per
vortex crossing the interface we use the expression
$I=j(\ell_{\perp}\ell_z)$, where $j=j_{V}$ from
Eq.~\eqref{eq:NeutronPhaseDiffCurrent} is the Josephson current and
$\ell_{\perp}$ and $\ell_{z}$ are the extent of the junction
perpendicular to/along the $z$ direction respectively. We find
\bea
\langle P_{V}\rangle
&=&\frac{(j_{0,pn}\omega_V)^2}{3c^3} (d\ell_{\perp} \ell_z)^2.
\label{eq:barP_n}
\eea
The radiant power from this mechanism for the entire star,
${\cal P}_{\star,V}$, can be estimated by approximating the spherical
geometry of the interface with an ``effective" cylinder of radius $R$
and height $\ell_z\sim R$, as explained above.  The number of vortices
on this cylinder at any time is determined by the ratio of its
circumference to the neutron intervortex distance, i.e.,
$N_V= 2\pi R/d_n$. Multiplying Eq.~\eqref{eq:barP_n} by the number of
vortex lines $N_V$, one finds
\bea 
\langle{\cal P}_{\star,V}\rangle &=&3\times 10^{7}~\tau_5^{-2}~R_6^{5}~\Omega_{100}^{3/2}~\textrm{erg~s}^{-1}, 
\label{eq:barP_nfinal}
\eea
where $\tau_5$ is the spin-down lifetime in units of $10^5$~yr, $R_6$ is
the stellar radius in units of $10^6$~cm, and and $\Omega_{100}$ is the
rotation angular frequency in units of 100 s$^{-1}$. We approximated
$d\sim\ell_{\perp}\sim\lambda$ and $\ell_z\sim R$. The details of the
computation are given in Appendix~\ref{app:B}.

The value in Eq.~\eqref{eq:barP_nfinal} is far too low to be
observationally interesting. For instance, it is far smaller than the
heating power due to the Ohmic dissipation of the crustal magnetic
field $B_{\rm cr}$ on a typical scale $\delta R$
\bea\label{eq:Ohmic}
{\cal P}_{\rm O}\simeq 10^{27}\left(\frac{B_{\rm cr}}{10^{13}~{\rm G}}\right)^2\left(\frac{10^{23}~{\rm s}^{-1}}{\sigma}\right)\left(\frac{\delta R}{0.5~{\rm km}}\right)\textrm{erg~s}^{-1},\nonumber\\
\label{eq:OhmicHeatingPower}
\eea
where $\sigma$ is crustal electric conductivity.

We now consider the radiation from a single proton flux tube passing through a
junction. Using $I=j_{\Phi}(\ell_{\perp}\ell_z)$ in Eq.~\eqref{eq:DipoleRadPower}, with $j_{\Phi}$
given by Eq.~\eqref{eq:p_current_t-dependent}, leads to
\bea
\langle P_{\Phi}\rangle
&=&\frac{2(j_0\omega_p)^2}{3c^3} (d\ell_{\perp} \ell_z)^2 \, 
\langle\cos^2(\Delta\theta_p + \phi(t)) \sin^2{(\omega_p
  t)}\rangle,\nonumber\\
&=&\frac{(j_0\omega_p)^2}{6c^3} (d\ell_{\perp} \ell_z)^2,
\label{eq:barP_p}
\eea
where we approximated the time average
\bea
&&\langle\cos^2(\Delta\theta_p+ \phi(t))\sin^2{(\omega_p t)}\rangle\nonumber\\
&&\hspace{1.5cm} = \frac{1}{4}\left\{1+J_1(2)\cos(2\Delta\theta_p)\right\}\simeq \frac{1}{4},
 \eea
where $J_1(x)$ is a Bessel function
and where we average over the possible phase differences
$\Delta\theta_p$.

According to our conjecture, the flux tubes are transported by
neutron vortex array over secular timescales with the average
velocity equal to the vortex line velocity Eq.~(\ref{eq:VortexLineSpeed}). 
The oscillation frequency of the time-dependent current $\omega_p$ is thus an average and depends
on the actual arrangement of the flux tubes in any particular model.
Therefore, the estimate of radiation power below should be viewed as an
estimate of the average output only. The details of the computation of the power of 
radiation are given in Appendix~\ref{app:B}. The final result is 
\bea
\langle P_{\star,\Phi}\rangle
&=& 2\times 10^{28}~ \tau_5^{-2}\, R_6^5~B_{13}^{3/2}~{\rm erg~s}^{-1},
\label{eq:barP_pfinal}
\eea
where $B_{13}$ is the magnetic field in units
$10^{13}$~G.

The value quoted in Eq.~\eqref{eq:barP_pfinal} can be compared to Eq.~(\ref{eq:OhmicHeatingPower}).
Unlike the heating caused by Josephson radiation linked to the time-dependent neutron phase from neutron vortex lines intersecting the $S$--$P$ interface, the heating due to the time-dependent proton phase from proton flux tubes crossing the $S$--$P$ interface is approximately an order of magnitude greater than the Ohmic dissipation \eqref{eq:Ohmic}, which is typically regarded as a primary heating mechanism in neutron stars.

\section{Conclusions}

In this work, we argued that the interface between $S$-wave and
$P$-wave superfluids can function like a Josephson junction, where a
supercurrent arises from the phase differences between the two
superfluids. We have shown that there is a stationary neutron current
across the interface due to this phase difference, which also entrains
part of the proton condensate thus generating a stationary flow of
protons across the interface.  Beyond this stationary effect, the
movement of neutron vortices and proton flux tubes generates
time-dependent Josephson currents, oscillating at a frequency
determined by the interflux tube spacing and neutron vortex
velocity.

A time-dependent electromagnetic current generates radiation. With
this in mind, we have demonstrated that the Josephson junction between
$S$- and $P$-superfluids produces electromagnetic radiation at the
interface, heating the star. Specifically, the amount of heat
generated by the time-dependent Josephson current induced by proton
flux tubes entering the junction is phenomenologically significant and
can impact the late-time cooling of neutron stars. The heating rate is
independent of temperature but depends on the spin frequency and its
derivative, as well as the strength of the magnetic field. Here, we
assumed a sharp interface between the $S$-wave and $P$-wave
superfluids. However, if the transition between these phases occurs
through a mixed phase, the radiation will be amplified due to the
``stacking” of multiple Josephson-like junctions along the path of a
flux tube cluster carried by a neutron vortex.  To confirm the
astrophysical importance of radiation from oscillating Josephson
currents, cooling simulations of neutron stars that incorporate the
discussed heating mechanism would be necessary. Such thermal cooling
simulations must also integrate the neutron star’s secular
deceleration dynamics under braking torques, such as magnetic dipole
radiation while accounting for the long-term evolution of the magnetic
field.

 We emphasize that the mechanism
  discussed here operates effectively on secular timescales and is
  not currently suitable for modeling transient phenomena in pulsar
  rotation, such as glitches. Moreover, its validity hinges on the
  assumption that, on average, flux tubes move together with neutron
  vortices. This assumption breaks down if neutron vortices are able
  to cut through flux tubes, leaving the latter stationary---a scenario
  that arises when significant restoring forces act on the flux tubes
  at their edges on the crust-core boundary.

 The dissipative process under consideration arises
  from the motion of neutron vortices, which reduces their number in
  the core. As a result, the rotational energy of the neutron
  superfluid serves as the ultimate energy source, with the associated
  radiation effectively acting as frictional dissipation at the
  boundary. This dissipation contributes to the internal torque
  transmitted from the superfluid core to the crust, which is
  incorporated into the coupled equations of motion for the two-fluid
  model with internal friction [see, for example,
  Ref.~\cite{Sedrakian1998ApJSpin}, Eqs.~(4) and (5)].

Finally, entrainment-induced nonstationary Josephson currents and
radiation proposed here may have analogs in other systems, for
example, in ultracold atomic systems, where Josephson junctions have
been recently manufactured~\cite{Pezze2024NatCo}. Entrainment would
require mixtures of atoms, whereas the electric current and radiation
would require a charged component or artificial gauge
fields~\cite{Galitski2019}. Such a setup is expected to work with any
combination of Fermi or Bose condensates, whether interfacing with
each other or forming mixtures.

\section*{Acknowledgments}

We thank Ira Wasserman for useful discussions. A.\,S. acknowledges 
support through Deutsche Forschungsgemeinschaft Grant
No. SE 1836/5-3, and the Polish NCN Grant No. 2020/37/B/ST9/01937. 
P.\,B.\,R. was supported by the Institute for Nuclear Theory's
U.S. Department of Energy Grant No. DE-FG02-00ER41132.

\begin{widetext}

\appendix
\section{Ginzburg-Landau model of superfluid neutron and
  superconducting proton fluids}
\label{app:A}

The Ginzburg-Landau theory provides a phenomenological framework
to describe fermionic superconductors and superfluids near their
critical points. Here, we describe a GL model for a proton superconductor and
a neutron superfluid, which extends the original GL theory to account
for coupling between the fluids. In the context of neutron stars such 
models were developed by various authors~\cite{Sedrakian1980,Alpar1984,Alford2008a,Wood2022}.

Consider a proton superconductor with an order parameter $\psi_p$ and
the neutron superfluid with an order parameter $\psi_n$. The general GL
free energy functional can be written as
\bea
F=F_0+F_{\rm  k i n}+F_{\rm  cond}+F_{\rm em},
\eea
where $F_0$ is the bulk free energy in the normal state,
$F_{\text {kin }}$ is the kinetic energy contribution which involves
the gradients of the order parameter, $F_{\text {cond}}$ is the
condensate term involving the order parameters, and $F_{\rm em}$ is the
electromagnetic field contribution. The energy is typically measured
from $F_0$; therefore, it can be dropped below. The condensation part
is given by
\bea
F_{\rm cond}= 
a_p\left|\psi_p\right|^2+\frac{b_p}{2}\left|\psi_p\right|^4+
a_n\left|\psi_n\right|^2+\frac{b_n}{2}\left|\psi_n\right|^4
+g_{np}\left|\psi_p\right|^2\left|\psi_n\right|^2,
\eea
where the coefficients $a_{p/n}$ and $b_{p/n}$ have their usual
meaning in the GL theory for
superfluidity/superconductivity~\citep{Tinkham1996}, and the term
$\propto g_{np}$ describes the coupling between the two
condensates. The part of the free-energy functional which represents
the kinetic energy contributions of two coupled fluids, with
velocities $\bm{v}_p$ and $\bm{v}_n$ is given by
\bea\label{eq:F_kin}
F_{\rm kin}=\frac{1}{2} \rho_{p p} v_p^2+\frac{1}{2}
\rho_{n n} v_n^2-\frac{1}{2} \rho_{p n}\left|\vec{v}_p-\vec{v}_n\right|^2,
\eea
where $\bm{v}_p$ and $\bm{v}_n$ refer to the velocities of the fluids
and $\rho_{p p}$ and $\rho_{n n}$ are the effective densities for the
superconductor and superfluid, respectively. $\rho_{p n}$ represents
the coupling between the two systems. These densities have been
derived using specific nuclear models~\cite{Borumand1996,Chamel2006a}
and can be expressed in terms of the effective masses of neutrons,
protons, and their partial densities $\rho_{n}$ and $\rho_{p}$.

In the GL theory for superconductors and superfluids, the velocities
$\bm{v}_p$ and $\bm{v}_n$ can be associated with the phase gradients
of the complex order parameters $\psi_p$ and $\psi_n$. Writing the
wave function with the magnitude and the phase
$\psi_j=\left|\psi_j\right| e^{i \theta_j}$ for $j=n,p$, the
velocities are given by the phase gradient of the order parameter,
plus the coupling to the vector potential for the protons,
\bea \label{eq:velocities}
\bm{v}_n=\frac{\hbar}{2m_n} \bm{\nabla} \theta_n, \qquad \bm{v}_p
=\frac{\hbar}{2 m_p}\left(\bm{\nabla} \theta_p-\frac{2e}{\hbar c} \bm{A}\right), \label{eq:SFVelocities}
\eea
where $\hbar$ is the Planck constant, $m_n$/$m_p$ is the mass of the
neutron/proton, and $\theta_n$/$\theta_p$ is the phase of the
superfluid neutron/superconducting proton order parameter. A factor $2
e$ appears instead of $e$ for the coupling of the protons to $\bm{A}$
because the superconducting protons form Cooper pairs with charge
$2e$.  Substituting Eqs.~(\ref{eq:SFVelocities}) into Eq.~(\ref{eq:F_kin}), we
obtain Eq.~\eqref{eq:F_kinMain} of the main text.

The velocity field of a quantized neutron vortex using the cylindrical coordinates $(\varrho,\varphi,z)$ is given by 
\bea \label{eq:vortex_v_n}
\bm{v}_n=\frac{n \hbar}{2m_n \varrho} \hat{\varphi}
\eea
where $n$ is the quantum number (in the ground state of lowest energy
$n=1$) $\varrho$ is the distance from the vortex center, $\hat\varphi$
is the angle variable with $z$ axis along the neutron vortex. For the
proton flux tube, the velocity field is given by
\bea\label{eq:vortex_v_p} \bm{v}_p=\frac{\Phi_0}{2 \pi \lambda}
K_1\left(\frac{\varrho}{\lambda}\right) \hat{\varphi},
\eea
where $K_1\left({\varrho}/{\lambda}\right)$ is the modified Bessel function of the second kind, which describes the radial decay of the velocity around the flux tube.

Finally, the electromagnetic field contribution to $F$ is
\bea
F_{\rm em} = \frac{(\bm{\nabla}\times\bm{A})^2}{8\pi}.
\eea
We assume that there is also an overall electrically neutralizing lepton fluid such that there are no electric fields in the problem.

The dynamics of ordinary GL theory is described by two equations of
motion that follow from the variation of the GL functional with
respect to the order parameter and electromagnetic field potential
$\bm{A}$, respectively. The number of equations of motion doubles in
the present case. The equations of motion for the proton and neutron
order parameters $\psi_p$ and $\psi_n$ are found by minimizing the
free energy with respect to $\psi_n^*$/$\psi_p^*$ respectively, which
leads to Eqs.~\eqref{eq:GL_p_1} and \eqref{eq:GL_n_1}.  These are the
coupled GL equations for the proton superconductor and neutron
superfluid, accounting for their interactions as well as the coupling
of the proton superconductor to the electromagnetic field.

To obtain the second set of  GL equations, which provides the proton charge current density and the neutron mass current density, we vary $F$ with respect to $\bm{A}$ and $\bm{v}_n$ respectively, giving
\bea
\bm{j}_p&=&\frac{e \hbar}{2 m_p^2 }\left(\rho_{p p}-\rho_{p
      n}\right)\left(\bm{\nabla} \theta_p-\frac{2 e}{\hbar c}
    \bm{A}\right)+\frac{e \hbar}{2 m_pm_n }\rho_{p n} \bm{\nabla} \theta_n,
    \label{eq:proton_current}
    \\
\bm{j}_n&=&\frac{\hbar}{2 m_n}\left(\rho_{n n}-\rho_{p n}\right)
    \bm{\nabla} \theta_n
    +\frac{\hbar}{2 m_p} \rho_{p n}\left(\bm{\nabla} \theta_p-\frac{2 e}{\hbar c} \bm{A}\right).
    \label{eq:neutron_current}
\eea
Unlike the ordinary currents in the GL theory, these
expressions depend on the phase gradients of both order parameters, coupled through the entrainment parameter $\rho_{np}$.

For the discussion of the junction, it is useful to eliminate the
phase gradients $\bm{\nabla}\theta_{p/n}$ from the expressions for the
currents. This is done using the relation
\bea\label{eq:rho-psi2}
\left(\psi_j^* \bm{\nabla} \psi_j-\psi_j\bm{\nabla}  \psi_j^*\right)
=2 i\left|\psi_j\right|^2\bm{\nabla} \theta_j= i\frac{\rho_{jj}}{m_j}\bm{\nabla} \theta_j , 
\eea
where we substituted $\psi_{j}=\left|\psi_{j}\right| e^{i\theta_{j}}$,
$j=n,p$ and defined $2m_j\left|\psi_j\right|^2 = \rho_{jj}$. Using
this relation, Eqs.~\eqref{eq:proton_current}
and~\eqref{eq:neutron_current} can be written in the forms
Eqs.~\eqref{eq:GL_p_2} and \eqref{eq:GL_n_2}, respectively.

\section{Computing the power of radiation}
\label{app:B}

We first show the computation of
Eq.~(\ref{eq:barP_nfinal}). Substituting the velocity of the vortex
lattice in the radiation formula we obtain
\eqref{eq:barP_n} 
\bea\label{eq:barP_n1}
\langle P_V\rangle
&=&\frac{(j_{0,pn}\omega_V)^2}{3c^3} (d\ell_{\perp} \ell_z)^2 =
\frac{j_{0,pn}^2}{3c^3}
\left(\frac{d\ell_{\perp} \ell_z}{d_n}\right)^2  \left(\frac{\dot \Omega}{2\Omega}\right)^2r^2.
\eea
To evaluate $j_{0,pn}$, we assume $|\psi_{1,n}|\approx|\psi_{2,n}|$,
because a density jump at the $S$--$P$ interface is not expected. We
also set $\rho_j=m_jn_j=2m_j^*|\psi_j|^2$, $j=n,p$, to define the bare
mass density in terms of the effective mass $m_j^*$ and $|\psi_j|^2$. We
thus find
\bea\label{eq:j_0}
j_{0,pn}=\frac{e\hbar}{m_p\zeta_n}\frac{\rho_{pn}}{\rho_{pp}}|\psi_{n}|^2
= e\eta_n v_{Fn} n_n,
 \qquad
\kappa_{np}\equiv \frac{\rho_{p n}}{\rho_{pp}},\qquad \eta_n
\equiv
\frac{\kappa_{np}}{2\zeta_nk_{Fn}},
\eea
where we introduced the neutron Fermi wave-vector $k_{Fn}$ and velocity
$v_{Fn}= \hbar k_{Fn}/m^*_n$. 

The problem involves two distinct characteristic scales: 1) a
microscale, on the order of 100 fm, which corresponds to the
penetration depth, $\lambda$, or several times the coherence length
, which represents the width of the Josephson junction, and 2) a macroscale,
on the order of the interface radius, $R$. We then approximate
\bea
d \simeq l_{\perp} \simeq \lambda, \quad l_z \simeq R.
\eea

To find the radiation from the entire interface (on the stellar
scale), we replace the spherical surface of the interface of two
superfluids with an ``effective cylinder'' and multiply the radiation formula \eqref{eq:barP_n1} by the number of vortex lines on this cylinder
$2\pi R/d_n$, i.e., this number is given by the circumference of the
cylinder divided by the intervortex line distance.  Thus, substituting
also the expression for the supercurrent amplitude \eqref{eq:j_0} we
obtain
\bea\label{eq:barP_n2}
\langle P_{\star,V}\rangle
&=& \frac{ 2\alpha_e\pi \hbar }{3c^2} \left(\frac{R}{d_n}\right)\left(\eta_n  v_{Fn} n_n \right)^2\,
\left(\frac{d\ell_{\perp} \ell_z}{d_n}\right)^2  \left(\frac{\dot
    \Omega}{2\Omega}\right)^2r^2\nonumber\\
&=& \frac{ \alpha_e\pi \eta_n^2}{6} \left(\frac{ v_{Fn} }{c}\right)^2
\left(n_n \lambda^2 R\right)^2 \left(\frac{R}{d_n}\right)^3 \frac{\hbar}{\tau^2},
\eea
where $\alpha_e = e^2/\hbar c \approx 1/137$ and in the second step we
introduced the pulsar lifetime $\tau^{-1} = |\dot\Omega|/\Omega$ and set the
location of the interface at $r=R$. To obtain the numerical estimate of the power of radiation we identify
the location of the transition density from $S$-wave to $P$-wave
superfluidity at the density of the crust-core interface
$n^{\sharp}= 0.08~\textrm{fm}^{-3}$. The radius of the cylindrical
interface is thus $R\approx 10^6$~cm.  

Next is the computation of Eq.~(\ref{eq:barP_pfinal}). Analogous
manipulations to those for the neutron vortex line-induced Josephson
current case result in
\bea\label{eq:barP_p2}
\langle P_{\star,\Phi}\rangle
&=& \frac{ \alpha_e\pi \eta_p^2}{12} \left(\frac{ v_{Fp} }{c}\right)^2
\left(n_p \lambda^2 R\right)^2 \left(\frac{R}{d_p}\right)^3 \frac{\hbar}{\tau^2},
\eea
where we have defined a proton Fermi velocity $v_{Fp}$ analogously to $v_{Fn}$, and where
\begin{equation}
    \eta_p\equiv\frac{1-\kappa_{np}}{2\zeta_p k_{Fp}}.
\end{equation}

To make further progress, we need to estimate the order of magnitude
of the microparameters of the proton and neutron condensate. For
fixed $ n^{\sharp}=0.08$ fm$^{-3}$, the effective mass values are
$m_p^*/m_p\approx 0.8$ and $m_n^*/m_n\approx 0.9$, as calculated at
this density in Ref.~\citep{Chamel2006a}. The neutron Cooper pairing gap at
this density is approximated by $\Delta_n=0.1$ MeV although
 variations can occur depending on the many-body
approximations involved~\citep{Sedrakian2019EPJA}. The neutron Fermi
wave number, Fermi velocity, and coherence length are thus
\bea
k_{Fn} &=& (3\pi^2 n_n)^{1/3} = 1.33~\textrm{fm}^{-1}, \quad \quad \frac{v_{Fn}}{c} =
\frac{\hbar k_{Fn}}{m_n^* c} = 0.28\left(\frac{m_n}{m_n^*}\right) 
= 0.31\left(\frac{0.9}{m_n^*/m_n}\right),
\eea
\bea
\xi_n (n_n^{\sharp})&=& \frac{\hbar^2c^2k_{Fn}}{\pi m_n^*c^2 \Delta_n}
=195~\left(\frac{k_{Fn}}{1.33~\textrm{fm}^{-1}}\right)
  \left(\frac{0.9}{m_n^*/m_n}\right) \left(\frac{0.1\ \textrm{MeV}}{\Delta_n}\right)~\textrm{fm}.
\eea
The neutron intervortex distance is
\bea
d_n = \left(\frac{\pi\hbar}{\sqrt{3}m_n^*\Omega}\right)^{1/2}
=3.56\times 10^{-3}\left(\frac{0.9}{m_n^*/m_n}\right)^{1/2}\left(
  \frac{100~\textrm{s}^{-1}}{\Omega}\right)^{1/2}~\textrm{cm}.
\eea
Finally, the parameter $\eta_n$ is computed, using $\kappa_{np}=m_p^*/m_p-1=-0.15$~\citep{Borumand1996} and $\zeta_n\approx 2\xi_n$,
\begin{equation}
    \eta_n=\frac{\kappa_{pn}}{2\zeta_n k_{Fn}}=-1.4\times10^{-4}\left(\frac{\kappa_{pn}}{-0.15}\right)\left(\frac{195\ {\rm fm}}{\xi_n}\right)\left(\frac{1.33\ {\rm fm}^{-1}}{k_{Fn}}\right).
\end{equation}

For the protons, we assume a proton fraction $1\%$ of total nucleonic
density at the interface (see Fig.~\ref{fig:gaps}). We then find the proton Fermi wave number and Fermi velocity
\bea k_{Fp} = (3\pi^2
n_p)^{1/3}=0.29\left(\frac{x_p}{0.01}\right)^{1/3}~\textrm{fm}^{-1}, \quad \quad \frac{v_{Fp}}{c} =
\frac{\hbar k_{Fp}}{m_p^* c} = 0.061\left(\frac{m_p}{m_p^*}\right)  = 0.076\left(\frac{0.8}{m_p^*/m_p}\right).
\eea
The proton coherence length and magnetic field penetration depth are
given by, using a proton Cooper pairing gap of $\Delta_p=0.1$ MeV,
\bea \xi_p (n_n^{\sharp})&=& \frac{\hbar^2c^2k_{Fp}}{\pi
  m_p^*c^2 \Delta_p} = 48
\left(\frac{k_{Fp}}{0.29\ \textrm{fm}^{-1}}\right)
\left(\frac{0.8}{m_p^*/m_p}\right) \left(\frac{0.1\ \textrm{MeV}}{\Delta_p}\right)~\textrm{fm},\\
\lambda (n_n^{\sharp})&=& \left( \frac{m_p^* c^2}{4\pi n_p \hbar
    c\alpha_e}\right)^{1/2} = 228 \left(\frac{m_p^*/m_p}{0.8}\right)^{1/2}
\left(\frac{0.01}{x_p}\right)^{1/2}~\textrm{fm}.  \eea
Using these parameters and $1-\kappa_{np}= 1.15$, which follows
from the value of $m_p^*/m_p=0.8$, and taking $\zeta_p\approx 2\xi_p$, we find
\begin{equation}
    \eta_p=\frac{1-\kappa_{pn}}{2\zeta_p k_{Fp}}=0.02\left(\frac{1-\kappa_{pn}}{1.15}\right)\left(\frac{48\ {\rm fm}}{\xi_p}\right)\left(\frac{0.29\ {\rm fm}^{-1}}{k_{Fp}}\right).
\end{equation}

We also need the interflux tube distance $d_p$,
which is obtained from flux density $n_\Phi$
\bea
n_\Phi = \frac{B}{\Phi_0} = 4.8\times 10^{19} \left(\frac{B}{10^{13}{\rm G}}\right),
\quad
d_p = \left(\frac{2}{\sqrt{3}n_\Phi}\right)^{1/2} = 1.55 \times 10^{-10}~{\rm cm},
\eea
where $B$ is the magnetic field induction. Substituting into
Eqs.~\eqref{eq:barP_n2} and \eqref{eq:barP_p2}, the parameters evaluated
above, we find Eqs.~\eqref{eq:barP_nfinal} and \eqref{eq:barP_pfinal} of
the main text.  The values of various dimensionless ratios entering
these expressions for the radiation are
\bea
\frac{\alpha_e\pi \eta_n^2}{6} = 7.49\times 10^{-11},\quad 
\left(\frac{ v_{Fn} }{c}\right)^2 =0.096,\quad 
\left(n_n \lambda^2 R\right)^2 = 1.73\times 10^{45} ,\quad 
\left(\frac{R}{d_n}\right)^3 =2.22\times 10^{25},
\eea
\bea
\frac{\alpha_e\pi \eta_p^2}{12} = 7.64\times 10^{-7},\quad 
\left(\frac{ v_{Fp} }{c}\right)^2 =0.0058,\quad 
\left(n_p \lambda^2 R\right)^2 = 1.72\times 10^{41} ,\quad 
\left(\frac{R}{d_p}\right)^3 =2.68\times 10^{47},
\eea
and finally, ${\hbar}/{\tau_5^2} = 10^{-52} \, {\rm erg}\ {\rm s}^{-1}$. 

\end{widetext}
\providecommand{\href}[2]{#2}\begingroup\raggedright\endgroup


\end{document}